\documentclass[aps,prl,floatfix,showpacs,twocolumn,footinbib]{revtex4}
\usepackage{epsfig}
\begin{document}


\title{Spin switch and spin amplifier: magnetic bipolar transistor 
in the saturation regime}

\author{Jaroslav Fabian} 
\affiliation{Institute for Theoretical Physics, Karl-Franzens University, 
Universit\"atsplatz 5, 8010 Graz, Austria} 

\author{Igor \v{Z}uti\'{c}}
\affiliation{Center for Computational Materials Science,
Naval Research Laboratory, Washington, D.C. 20375 and
Condensed Matter Theory Center, Department of Physics, 
University of Maryland at College
Park, College Park, Maryland 20742-4111}

\begin{abstract}
It is shown that magnetic bipolar transistors (MBT) can amplify currents 
even in the saturation regime, in which both the emitter-base and
collector-base junctions are forward biased. The collector current and the 
current gain can change sign as they depend on the  relative
orientation of the  equilibrium spin
in the base and on the nonequilibrium spin 
in the emitter and collector. The predicted phenomena should be
useful for electrical detection of nonequilibrium spins in semiconductors,
as well as for magnetic control of current amplification and for current
switching.
\end{abstract}
\maketitle

\section{Introduction}

Silsbee's 
prediction \cite{Silsbee1980:BMR} that nonequilibrium spin at an interface
between normal and ferromagnetic metals gives rise to 
electromotive force (emf)
was verified in the original Johnson-Silsbee spin injection 
experiment \cite{Johnson1985:PRL} with aluminum and permalloy. The observed voltages
across the interface were on the order of picovolts, registering
about one per $10^{11}$ spins in aluminum. An interesting application
of the coupling was suggested by Johnson under the name of
bipolar spin transistor \cite{Johnson1995b:JMMM} (polarity 
there refers to spin, not charge), 
or spin switch \cite{Johnson1993:S}, an all metallic structure. 
In that 
proposal the current in the collector changes sign depending
on the relative orientation of the two ferromagnets forming the switch. The
scheme is explained in Fig. 1. Being all metallic, the device offers
no current gain.

\begin{figure}
\centerline{\psfig{file=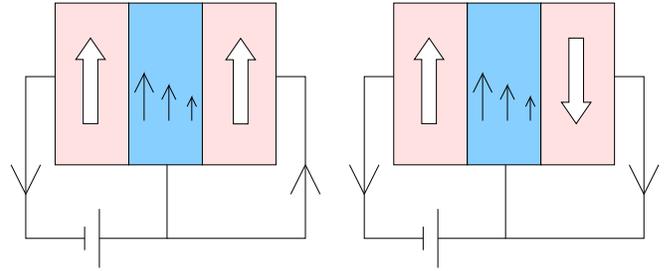,width=1\linewidth}}
\caption{Johnson's spin switch. In the metallic
trilayer structure the two outer layers are ferromagnetic. 
Spin-polarized 
electrons are emitted from the left layer
to the middle nonmagnetic metal. Assuming that spin polarization
survives the diffusion to the right layer, the electrons
either enter the right ferromagnet, if the magnetizations
are parallel, or bounce back if the magnetizations are
antiparallel. The spin-dependent emf appearing at the
interface between the middle and right layers is due to 
the imbalance of the spin-resolved chemical potentials. 
The current in the right ferromagnet (collector) can be 
switched by changing the relative magnetization
orientation of the two ferromagnets.
}
\label{fig:scheme_eq}
\end{figure}

We have shown that spin-charge coupling is significantly more
pronounced in inhomogeneous magnetic semiconductors at nondegenerate
doping levels \cite{Zutic2002:PRL, Fabian2002:PRB}. 
By magnetic we mean either ferromagnetic
(though these may be hard to make nondegenerate) or large-Zeeman
effect (paramagnetic) semiconductors. The spin-charge coupling
increases exponentially with applied bias, leading to, for
example, giant magnetoresistance \cite{Zutic2002:PRL} (more than about 10\% change
in the relative resistance in a magnetic multilayer structure
with alternating magnetic and nonmagnetic layers, 
upon changing from parallel to antiparallel the magnetization
of the magnetic layers). 

Magnetic semiconductors integrated within the conventional device
schemes show great technological potential. 
Several device schemes have been suggested that use magnetic semiconductors
both as passive or active medium \cite{Zutic2004:RMP}. We have proposed to integrate
magnetic transistors in the trilayer structure of a bipolar transistor,
creating what we termed magnetic bipolar transistor \cite{Fabian2002:P} (MBT), in 
which any of the three regions (emitter, base, or collector)
can be magnetic. In addition, nonequilibrium spin is assumed
inside  the structure, giving rise to spin-charge coupling. 
We have shown how in the active forward (the usual mode
of operation) regime the current gain depends on the spin-charge
coupling, leading to what we called giant magnetoamplification
effect, which is a sensitive modulation of the charge current
amplification by magnetic fields and nonequilibrium spin.

In this paper we describe other peculiar phenomena of MBT's:
spin induced current gain and current switching (MBT as
a Johnson's spin switch) in the
saturation regime, which is usually the ON mode of the transistor
in logic circuits. Conventional transistors, in contrast, 
show little current gain in this regime.

\section{\label{sec:model} Magnetic bipolar transistor}

\begin{figure}
\centerline{\psfig{file=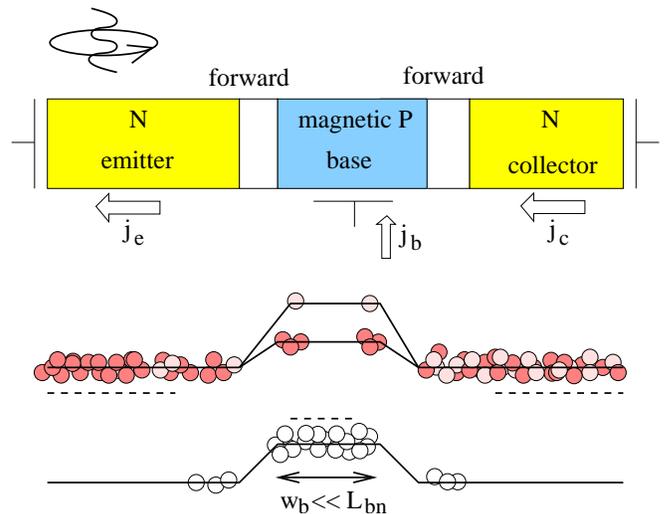,width=1\linewidth}}
\caption{Magnetic bipolar transistor. The upper figure shows the scheme
used in the text.
The emitter and collector are nonmagnetic n-doped
regions, while the base is magnetic p-doped. The emitter can have a
nonequilibrium spin injected optically or electrically. The currents
flowing through different regions are identified. The lower figure
shows the band diagram. The lower band is valence band, 
populated with holes (empty circles)  
which 
are assumed unpolarized. The upper 
band is the conductance band, 
with spin up (dark circles) 
and spin down (light circles)  
electrons. The conductance band is spin split in the base, indicating
equilibrium spin polarization. In the saturation regime there is
a forward bias (here taken equal) at both junctions. In the conventional
case this means that the collector and emitter electron currents
are the same, indicated by the equal drop of the chemical potential
(dashed lines). 
In MBT's the chemical potential depends on spin, giving
rise to spin-charge coupling.
}
\label{fig:scheme}
\end{figure}

The structure of a MBT is shown in Fig. \ref{fig:scheme}. The transistor
has three regions: emitter, base, and collector. In the scheme
used here the emitter and collector are nonmagnetic, doped with
donors (n-doped), while the base is magnetic, doped with acceptors
(p-doped). By magnetic we mean that the carrier bands are spin split. 
For simplicity we consider only the conduction band to be spin 
split, leading to an equilibrium spin polarization of electrons.
Holes are assumed unpolarized. The spin splitting can develop
if the semiconductor is ferromagnetic, in which case the splitting
is due to exchange coupling, or the semiconductor 
is doped with magnetic impurities and placed in a magnetic field, in 
which case the splitting arises from the large Zeeman effect 
\cite{Zutic2004:RMP}.
For our purposes the origin of the splitting is not crucial,
we describe what happens if the splitting occurs.  
One of the implications of such splitting is a finite
spin polarization of the electron density
$\alpha=(n_\uparrow-n_\downarrow)/(n_\uparrow+n_\downarrow)$,
expressed in terms of the spin-resolved electron densities,
$n_\uparrow$, $n_\downarrow$.
We denote
the equilibrium spin polarization 
in the base as $\alpha_{0b}$, where the subscript "0" represents
an equilibrium quantity. 
An important aspect of MBT is the possibility of having
nonequilibrium spin in otherwise nonmagnetic regions. Spin
can be introduced into the emitter or collector either
optically \cite{Meier:1984} or electrically 
\cite{Fiederling1999:N, Jonker2000:PRB, Young2002:APL, Jiang2003:PRL}. 
We denote the nonequilibrium
(or excess) spin polarizations
in the two regions by $\delta \alpha_{e}$ and
$\delta \alpha_{c}$, respectively.
An  analytic theory of magnetic bipolar transistors was developed in 
Ref.~\cite{Fabian2004:PRB} by generalizing Shockley's model 
\cite{Shockley:1950}. 

We have previously described the physics of MBT's in the active
forward regime \cite{Fabian2002:P, Fabian2004:APL, Fabian2004:PRB}, 
in which the conventional bipolar transistor
exhibits current gain. We have shown that the transistor exhibits
magnetoamplification and giant magnetoamplification. The former arises
from the trivial change of the equilibrium carrier density 
with the change in the carrier band splitting 
(see also Ref.~\cite{Lebedeva2003:JAP}),
while the latter appears due to spin-charge coupling. In the
next section we describe the workings of MBT's in what is 
called the saturation regime, in which current gain in conventional
transistors is limited, but in MBT's it can be large, with
magnitudes as in the active forward regime. This
effect is solely due to spin-charge coupling. In order to understand
the predicted phenomena, we introduce below equations for the
currents flowing through the different regions (see Fig. \ref{fig:scheme}),
and their relations to the excess (nonequilibrium) carrier 
densities. We then use simplifying assumptions to derive
the currents and the current gain in the saturation regime. We
will also present a numerical calculation based on the full
analytical theory with no further simplifications.

The charge currents flowing in the emitter ($j_e$), collector ($j_c$), and base 
($j_b$) can be written in the form usual for nonmagnetic bipolar transistors
\cite{Fabian2004:PRB}:
\begin{eqnarray} \label{eq:je}
j_e&=&j_{gb}^n\left [\frac{\delta n_{be}}{n_{0b}}-\frac{1}{\cosh(w_b/L_{nb})}
\frac{\delta n_{bc}}{n_{0b}} \right ] + j_{ge}^p\frac{\delta p_{eb}}{p_{0e}},\\
\label{eq:jc}
j_c&=&j_{gb}^n\left [-\frac{\delta n_{bc}}{n_{0b}}+\frac{1}{\cosh(w_b/L_{nb})}
\frac{\delta n_{be}}{n_{0b}} \right ]- j_{gc}^p\frac{\delta p_{cb}}{p_{0c}}.
\end{eqnarray}
The base current (for the signs of the currents see Fig. \ref{fig:scheme} is  
\begin{equation}
j_b=j_e-j_c.
\end{equation}
The equilibrium electron density in the base is $n_{0b}$, while the
excess density 
in the base close to the junction with the emitter (collector) 
is $\delta n_{be}$ ($\delta n_{bc}$). Similar notation is used for the 
hole densities $p$. The electron diffusion length through the base of
width $w_{b}$ is $L_{nb}$. 
The electron contributions to the currents are proportional to the electron
generation current in the base, which is
\begin{equation} \label{eq:jg}
j_{gb}^n=\frac{qD_{nb}}{L_{nb}}n_{0b}\coth\left(\frac{w_b}{L_{nb}}\right).
\end{equation}
Here $D_{nb}$ stands for the electron diffusion coefficient in the base.
Similarly, the hole contribution depends on the respective hole generation
currents in the emitter ($j_{ge}^p$) and collector ($j_{ge}^p$). These two
generation currents have expressions analogous to Eq.~\ref{eq:jg}.

The difference between the conventional and magnetic transistor appears
in the expressions for the nonequilibrium carrier densities. The excess
electron density in the base close to the emitter 
($\delta n_{be}$) and collector ($\delta n_{bc}$) 
are respectively \cite{Fabian2004:PRB}
\begin{eqnarray}
\delta n_{be} &=&n_{0b} e^{qV_{be}/k_B T} 
\left (1+\alpha_{0b}\delta \alpha_{e}  \right ), \\
\delta n_{bc} &=&n_{0b} e^{qV_{bc}/k_B T} 
\left (1+\alpha_{0b}\delta \alpha_{c}  \right ),
\end{eqnarray}
which follow from generalizing the implications
of the spin-voltaic effect (a particular form of a spin-charge
coupling), studied previously in magnetic diodes 
\cite{Zutic2002:PRL,Fabian2002:PRB}.
In conventional transistors the excess densities are determined solely by the Boltzmann
factors containing the base-emitter and base-collector biases
$V_{be}$ and $V_{bc}$, respectively, and temperature $T$. In MBT's the excess densities
depend also on the spin-charge factors containing the equilibrium
spin polarization in the base $\alpha_{0b}$ and nonequilibrium 
spin polarization in the emitter and collector, $\delta \alpha_e$ and
$\delta \alpha_c$.  
For no spin-charge coupling the above equations reduce to the well known 
formulas for electron injection through a depletion layer \cite{Shockley:1950}. Since 
holes are unpolarized, their excess densities in the emitter
and collector are
\begin{eqnarray}
\delta p_{e} &=& p_{0e} e^{qV_{be}/k_B T}, \\
\delta p_{c} &=& p_{0c} e^{qV_{bc}/k_B T}. 
\end{eqnarray}
In what follows we apply Eqs.~\ref{eq:je} and \ref{eq:jc}   to 
a particular case of MBT in the saturation regime.

\section{\label{sec:saturation}  Magnetic bipolar transistor in the saturation regime}

In the saturation regime both the emitter-base and base-collector junctions are
forward biased. We consider first a special case of equal forward biases, 
$V\equiv V_{be}=V_{bc} \gg k_B T$. We also assume that the emitter and collector
doping is larger than the base one, so that $p_{0e}, p_{0c} \ll n_{0b}$ 
(the equilibrium minority carrier density is inversely proportional to the
doping density)
and we will be able to neglect the hole contributions to the currents.
Such doping is unusual, since, due to technological reasons, in conventional 
bipolar transistors the collector doping is small.  
We note that the
normal (active forward) regime of transistor operation is a forward emitter-base
bias $(V_{be} > 0)$ and a reverse base-collector bias $(V_{bc} < 0)$, to 
ensure amplification. 

\begin{figure}
\centerline{\psfig{file=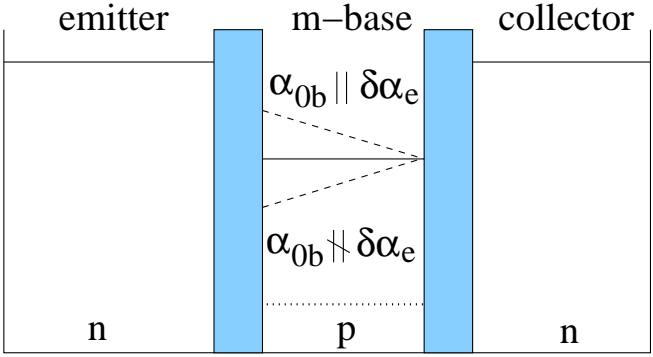,width=1\linewidth}}
\caption{Spatial profile of the electron density in an MBT in the saturation regime.
In the n-doped emitter and collector the density is almost constant,
at the donor doping level to ensure neutrality. In the magnetic base (m-base), the equilibrium
density (dotted line) 
is very small, given by 
$n_{0b}=n_i^2/N_{ab}\cosh(q\zeta_b/k_B T)$, where $n_i$ is
the intrinsic carrier density at a given temperature and $N_{ab}$ is
the acceptor doping in the base; $2q \zeta_b$ is the conduction band 
splitting in the base. If forward bias $V$ is applied, the
electron density is raised to $n_{0b}\exp(qV/k_B T)$, shown by the
solid line. If the same forward
bias is applied to the emitter-base and collector-base junction,
the electron density in the base will be almost uniform, giving
rise to only small diffusion currents in the emitter and collector.
If, however, the emitter spin is out of equilibrium, the injected
electron density in the base will be $\delta n_{be}=n_{0b}\exp(qV/k_B T) 
(1+ \alpha_{0b}\delta \alpha_e)$,
allowing an increase or decrease of $n$ depending on the relative orientation
of $\alpha_{0b}$ and $\delta \alpha_{e}$ (dashed lines for parallel and antiparallel
orientations). Significant current can flow in the collector.  
}
\label{fig:profile}
\end{figure}

Figure \ref{fig:profile} illustrates the spatial profile of the electron
density in a MBT. The excess electron density in the base determines
how much current flows in the emitter and collector. If no spin-charge
coupling is present, the density is almost uniform, creating only
small diffusive currents. However, by changing the density at
one junction by turning on spin polarizations, the density
gradient gets larger leading to large collector and emitter currents.
This is how spin induces current in MBT's. On the other hand, since
the contributions from the spin-charge couplings to $j_e$ and $j_c$
are similar, they cancel in the base current $j_b=j_e-j_c$. This is
how the base current remains small, leading to large spin-induced
current gain for small signal amplifications.  

Take the collector current first. From Eq.~\ref{eq:jc}. Substituting for
the nonequilibrium electron density, the following expression is obtained in the
narrow base approximation ($w_b \ll L_{nb}$):
\begin{equation} \label{eq:jc1}
j_c=j_{gb}^n e^{qV/k_B T} \left [\alpha_{0b}\left (\delta \alpha_e - 
\delta \alpha_c  \right ) - \frac{w_b^2}{2L_{nb}^2}  - 
\frac{j^p_{gc}}{j^n_{gb}}\right ] .
\end{equation}  
If either the equilibrium or the nonequilibrium polarization vanishes,
the charge current flowing through the collector is of the order of
$j_g \exp(qV/k_B T) max(w^2/L^2, j^p_{gc}/j^n_{gb})$. This current
is small in the narrow-base limit and in the limit of large
collector doping. If a significant 
spin polarization is present, the
current is much greater, on the order of $j_g \exp(qV/k_B T)$. Similar
considerations hold for the emitter current.

Assuming that $\alpha_{0b} \delta\alpha_{e(c)} >> w^2/L^2$, $j^p_{gc}/j^n_{gb}$, the important
qualitative effect of Eq.~\ref{eq:jc1} is that 
\begin{equation} \label{eq:jcq}
j_{c} \sim \alpha_{0b} (\delta\alpha_{e}
-\delta\alpha_{c}).      
\end{equation}
Suppose there is a nonequilibrium spin either in the
emitter or in the collector (or in both). 
The sign of the collector current
can be changed by switching the magnetic field (changing $\alpha_{0b}$, or
changing the sign of the nonequilibrium spin polarizations). Since $j_{gb}^n$
is usually known, measurements of $j_c$ can directly give the product
$\alpha_0\delta \alpha_{e(c)}$ and thus directly detect the presence
of nonequilibrium spin. On the other hand, the transistor functions as
Johnson's spin switch: by changing the relative orientation 
of the equilibrium and nonequilibrium spins, $j_c$ changes sign.  
  
Another marked difference from the conventional bipolar transistor in the
saturation regime is the emergence of a current gain. Current gain $\beta$,
defined as
\begin{equation}
\beta = \frac{j_c}{j_b},
\end{equation}
measures the amplification efficiency of the transistor. In the active 
forward regime $\beta$ is typically 100. In the saturation regime $\beta$
becomes small, on the order of one. In MBT's, however, $\beta$ is large
even in the saturation regime. Indeed, using our assumptions of magnetic
and narrow base, and large emitter and collector dopings, the current 
gain is
\begin{equation} \label{eq:beta}
\beta=\frac{\alpha_{0b}\left 
( \delta \alpha_e - \delta \alpha_c \right )}{w_b^2/L_{nb}^2+ 
j_{ge}^p/j_{gb}^n + j_{gc}^p/j_{gb}^n}.
\end{equation}
The current gain is large because the prefactor, the ratio of the electron
and hole generation currents is large, $\sim n_{0b}/p_{0e(c)}$. The
gain becomes small (on the order of one)
if either of the spin polarizations
vanishes. The remarkable fact is that in the saturation regime
the current gain can even be negative.

\begin{figure}
\centerline{\psfig{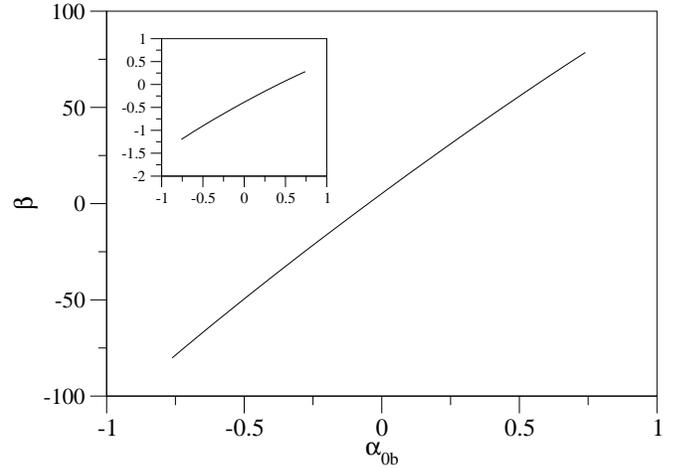}}
\caption{Spin-induced 
current gain of MBT's as a function of 
the equilibrium spin polarization in the base
$\alpha_{0b}$ 
in the saturation regime. The nonequilibrium spin polarization is $\delta\alpha_e=10\%$
($\delta\alpha_c=0$). 
The main graph
is for Si-like materials parameters, the inset is for GaAs-like materials parameters.
The geometry is the same in both cases. Gain is less significant in the GaAs-like
case, due to the breakdown of the narrow base assumption. 
 The cure is to make the
The remedy would be to make the
base narrower.
}
\label{fig:beta}
\end{figure}

The above approximations should be a robust description of the actual
behavior in MBT's. Using our analytic theory \cite{Fabian2004:PRB} 
for ideal MBT's, without making any further assumptions, 
we show $\beta=\beta(\alpha_{0b})$
in Fig. \ref{fig:beta} for a realistic MBT with generic 
room temperature materials
parameters, one set borrowed from Si (main graph), one from GaAs (inset).
The geometry and the doping profiles are the same. The emitter and the base have nominal lengths
of 2 $\mu$m, the base is 1 $\mu$m. The emitter and collector donor densities
are $10^{17}$/cm$^3$, the base acceptor density is $10^{16}$/cm$^3$. The
applied forward voltages are 0.5 volts. Other parameters, 
such as diffusion coefficients 
and spin relaxation times,
are given in Ref.~\cite{Fabian2004:PRB}. The
narrow base approximation is well valid for the Si-like case, where
$L_{nb}\approx 30$ $\mu$m, while it is not generally valid for our GaAs-like case,
in which $L_{nb}\approx 3$ $\mu$m. The results for Si show efficient 
spin control of the current gain, with the magnitude of $\beta$ similar
to the normal active regime. In GaAs the gain is small, with the magnitude
as in conventional bipolar transistors, although the sign of $\beta$
is still changed upon changing $\alpha_{0b}$. The results of the
full analytic theory are consistent with the approximate formula
Eq.~\ref{eq:beta}.

Finally, we ask what happens if the biases are unequal, 
$V_{be}\ne V_{bc}$. Suppose
that $V_{bc}=V_{be}+\Delta V$. The above effects are valid  $q\Delta V \ll k_B T$
(perhaps if $q\Delta V \le 0.1 k_B T$). However, one can tune $\Delta V$ in 
order to maximize the spin effects. It is straightforward to show, for example, that
in the narrow-base approximation if 
\begin{equation}
e^{-q\Delta V/k_B T} = 1+\frac{w_b^2}{2L_{nb}^2} + \frac{j_{gc}^p}{j_{gb}^n},
\end{equation}
the collector current obeys Eq.~\ref{eq:jcq}. This relaxes
the assumption
of large collector doping and makes the predictions more robust. However,
even if the spin control of $j_c$ is not possible because the base is
wide or the collector and acceptor doping are small, the spin effects
(and thus spin injection and detection) 
can be
observed in quantities
such as 
$j_c(\alpha_{0b})-j_c(-\alpha_{0b})$, where the 
 spin-independent 
parts 
cancel out.

\section{Conclusions}

We have described a novel phenomenon of spin-induced
current gain in 
magnetic bipolar transistors in the saturation regime with similar
emitter-base and base-collector forward bias voltages. Further crucial
assumptions include narrow base, to make carrier diffusion through
the base efficient, and small base acceptor doping, to neglect the
hole currents. The transistor's collector and emitter currents
can be switched by changing the relative orientation of the 
equilibrium spin polarization in the base and the nonequilibrium
spin polarization in either the emitter or the collector. The gain
too can change sign. In contrast to the conventional bipolar
transistor, the current gain can be as large as in the active forward
regime. These effects should be useful for sensing magnetic fields
and for the spin control of current amplification. They also allow for
measurements and detection of nonequilibrium spin densities in 
semiconductors, which should be useful especially in semiconductors
in which optical spin detection is ineffective.  
There has been steady experimental progress in recent years towards bipolar 
(with respect to charge) spintronics 
\cite{Karczewski1981:SSC,Janik1988:APPA,Kohda2001:JJAP,Johnston-Halperin2002:PRB,Ohno2000:ASS,%
Tsui2003:APL,Zutic2004:RMP},
which together with the exciting theoretical predictions warrants
future investigations of the subject.

\acknowledgments{We are grateful to S. Das Sarma for useful discussions. 
I. \v{Z}. acknowledges the National Research Council for financial support. 
This work was funded by DARPA, the NSF-ECS, and the US ONR.

\bibliographystyle{apsrev}
\bibliography{references}

\end{document}